\documentstyle[12pt]{article}
\begin{document} 
\thispagestyle{empty} 
\begin{flushright}
UA/NPPS-3-02\\
hep-ph/0205296\\
\end{flushright}
\begin{center}
{\large{\bf
A GLOBAL ASSESSMENT OF THE STRANGENESS-INCLUDING STATISTICAL BOOTSTRAP MODEL
ANALYSIS OF NUCLEUS-NUCLEUS AND $\bf p-\bar{p}$ INTERACTIONS\\}}
\vspace{2cm} 
{\large A. S. Kapoyannis, C. N. Ktorides and A. D. Panagiotou}\\ 
\smallskip 
{\it University of Athens, Division of Nuclear and Particle Physics,\\ 
GR-15771 Athens, Greece}\\ 
\vspace{1cm}

\end{center}
\vspace{0.5cm}
\begin{abstract}
A strangeness and isospin asymmetry including statistical bootstrap model
analysis of the multiparticle system produced in the $Pb+Pb$ collision at
158 AGeV at CERN is presented. It is concluded that this interaction
process has not crossed the deconfinenent line. Direct comparisons with the
results of similar analyses pertaining to nucleus-nucleus and $p-\bar{p}$
collisions at CERN are made. The overall picture points to the $S+S$ collision
at 200 AGeV as a prime candidate of a process which has crossed the border
separating the hadronic from the deconfined phase of matter.
\end{abstract}

\vspace{3cm}

e-mail addresses: akapog@cc.uoa.gr, cktorid@cc.uoa.gr, apanagio@phys.uoa.gr

PACS numbers: 25.75.-q, 12.40.Ee, 12.38.Mh, 05.70.Ce

Keywords: Statistical Bootstrap Model, Heavy Ion Collisions, Deconfined Phase

\newpage
%\vspace{2cm}
\setcounter{page}{1}

The quintessential issue probed by relativistic heavy ion experiments
concerns whether or not a phase of matter beyond the hadronic one has been
achieved during the interaction process. Any conclusive evidence, furnishing
an affirmative answer to this question, is relevant not only in solidifying
the status of QCD, as the microscopic theory for the strong interaction, but
also in providing a short glimpse into the early cosmos.

The most tangible element of experimental input, on which one relies to
infer as to what has taken place during a given collision between two heavy
nuclei, is the constitution of the produced multiparticle system. One
viewpoint to examine is whether such a system originates from a thermally
and chemically equilibrated state, having gone through a complex stage of
dynamical evolution. This is the premise upon which a number of extensive
analyses have been conducted [1-3], pertaining to multihadron systems in heavy
ion collisions, as well as $p-\bar{p}$ and $e^+-e^-$. A common feature in the
aforementioned studies is the adoption of an ``ideal hadron gas'' model to
simulate such multiparticle systems. These kinds of approaches, however, can
go only so far as to verify the thermalization hypothesis - which, in fact,
they do - but fall short of addressing the basic question whether a given
observed multiparticle system provides any signature which could trace its
source inside or outside the hadronic phase of matter. This is due to the
absence of {\it interactions} in the thermodynamic description of such
relativistic, many-particle system in this type of calculations\footnote{Only
an excluded volume correction has been addressed in some of these studies.}.

The fundamental issue of constructing a self-consistent scheme, which
furnishes a thermodynamic account of an interacting relativistic multihadron
system was formulated in the 1960's by Hagedorn [4-7] and is known
as ``Statistical Bootstrap Model'' (SBM). In this scheme, the notion of
interaction at a distance between
relativistic particles is simulated by a successive organization of the
system in terms of particle-like entities of increasing complexity, known as
fireballs. At the lowest level enter all known hadrons, which are incorporated
as input into a self-consistent (bootstrap) iterative scheme, realized in
terms of an integral equation for the fireball mass spectrum function
$\rho(m)$. The final construction of the SBM is accomplished by enriching the
scheme, in a relativistically consistent way, via the introduction of
thermodynamic variables (temperature and fugacities), which monitor its
thermal and chemical equilibrium. After some interim mathematical
manipulations, the bootstrap equation acquires the form
\begin{equation}
\varphi(T,\{\lambda\})\;=\;2G(T,\{\lambda\})-\exp[G(T,\{\lambda\})]+1\;,
\end{equation}
where $\varphi$ is the so-called input function, whose specification is
determined by masses and quantum numbers of the known hadrons, while
$G$ incorporates the fireball mass spectrum. The most notable feature of the
bootstrap equation is that it displays, in the $\varphi-G$ plane, a
square-root branch point at
\begin{equation}
\varphi(T_{cr},\{\lambda_{cr}\})=\ln 4-1\;,
\hspace{2cm} G(T_{cr},\{\lambda_{cr}\})=\ln 2\;. 
\end{equation}
The physical branch serves to define a critical hyper-surface in the space of
thermodynamic variables, which sets the limits of the hadronic phase. Finally,
the ``statistical'' aspect of the SBM is encoded in a grand-partition function
$Z_{SBM}$. It is of crucial importance to note that the mass spectrum
function, which incorporates the interaction dynamics among the hadrons,
enters the expression for $Z_{SBM}$.

The task of employing the SBM to perform non-trivial (in the sense of
including interactions) thermal-type analysis of multiparticle yields from
heavy ion collisions has been undertaken by the present authors in recent
publications [8-11]. To this end, we extended the original Hagedorn scheme, in
which a single fugacity variable for baryon number was employed, so as to
include fugacities relating to strangeness numbers, both net (strangeness
minus anti-strangeness) $\lambda_s$ and absolute (strangeness plus
anti-strangeness) $\lambda_{|S|}$ [9-10], as well as a ``net charge''
fugacity [11]. The latter, in combination with the baryon number fugacity,
accounts for isospin asymmetry in the colliding system. We formulated in this
way the Strangeness-including SBM (SSBM). Moreover, we departed [8-9] from
the usual choice of a certain parameter $\alpha$
entering the bootstrap scheme, which determines the particular partitioning
between a kinematical and a dynamical factor composing the mass spectrum
function $G$. In this way, a direct relationship arises between the maximum
value of the temperature, $T_0$, on the critical surface and the MIT bag
constant. The SSBM is analytically applicable only within and up to the limit of
the hadronic phase, defining in a precise way this limit. For our purposes
the boundary of the hadronic domain will be designated by its projection on
the 2-dimensional $(T,\mu_u)$ phase diagramme. This projection is close to
the intersection of the critical surface with the $\mu_s=0$ plane.

Given the wide interest generated by the $Pb+Pb$ experiments at SPS, we shall
present, in this letter, results stemming from an SSBM analysis of data at
158 AGeV. We shall also make a global assessment of all results we have so
far obtained via the SSBM approach, which include $S+S$, $S+Ag$ and
$p+\bar{p}$ collisions at SPS. Such an analysis is provided by the space of
thermodynamical variables, a comprehensive form of which is represented by
the set $(V,T,\lambda_u,\lambda_d,\lambda_s,\gamma_s)$, where
$\gamma_s=\lambda_{|S|}$. Two constraints are imposed on these variables,
corresponding to zero strangeness and the connection between total baryon
number and total charge of the colliding system
\begin{equation}
<S>=0,\hspace{2cm}\frac{<B>}{<Q>}=\frac{N_p^{in}+N_n^{in}}{N_p^{in}}\;, 
\end{equation}
where $N_p^{in}$ ($N_n^{in}$) is the total number of participant protons
(neutrons).

Referring to the first column of Table 1, we consider the full phase space
yields of particles produced in the Pb+Pb NA49-experiment at CERN [12,13] and
perform a $\chi^2$-fit in relation to the theoretical expressions furnishing
the corresponding particle multiplicities in the SSBM
\begin{equation}
N_i^{th}=\left.
\lambda_i\frac{\partial \ln Z_{SSBM}}{\partial\lambda_i}
\right|_{\{\lambda_i=1\}}
\end{equation}
through which an optimum value for each one of the thermodynamical variables
is determined\footnote{Actually, along with the thermodynamical
variables, two additional ones enter the $\chi^2$-fit, namely the two
Lagrange multipliers that serve to impose the constraints.}[9].

Columns 3 and 4 of Table 1 present the SSBM calculated particle yields
corresponding to optimum values of the thermodynamical variables, obtained
with and without the inclusion of pions, respectively. The corresponding
optimum values of these variables are entered in Table 2. Finally, in Table 3
we list the experimentally measured particle ratios, upon which we shall base
the presentation of our results. In Fig. 1 we depict the location of the
multiparticle emitting source on the $T-\mu_u$ plane, resulting from the SSBM
analysis with the inclusion of pion multiplicities and the fugacity variable
$\gamma_s$ fixed to its optimal value 0.76 (see second column of Table 2).
For each particle ratio we have
formed corresponding bands, as determined by experimental uncertainties. The
source coordinates, obtained from the fit to the particle yields, are located
at the point which is nearer to all the particle-ratio bands, identifying the
region of the original achievement of thermal and chemical equilibrium by the
multiparticle system. The most significant feature in the plot is the thick
solid line, the SSBM (de)confinement line for the value of $\gamma_s$
corresponding to the $Pb+Pb$ interaction. It is an indigenous characteristic of the
SSBM and its particular specification in the plot involves the choice for the
critical temperature, $T_0$, at zero chemical potentials $\mu_u$, $\mu_d$ and
$\mu_s$ ($\gamma_s$ is fixed). The reasoning behind our particular choice:
$T_0\simeq 183$ MeV, corresponding to maximum MIT bag constant value
$B^{1/4}=235$ MeV [12] is that it allows as much as possible space to the
hadronic phase [9]\footnote{Also $T_0(\mu_q=0)\simeq 183$ MeV is the maximum
temperature for non-negative $\mu_s$ in the HG domain. Recent lattice QCD
calculations give $T_0\simeq 175$ MeV for [2+1] quark flavours [13].}. Thus
no doubt can be left that an interaction point is outside the
hadronic domain if it lies beyond the surface for which the hadronic domain
is maximally extended.

Let us point out that the recent data [14] give smaller values for the $4\pi$
multiplicities of $\Xi^-$, $\bar{\Xi}^+$ and for the $\bar{\Lambda}/\Lambda$
ratio. This leads to the lowering of the fitted temperature as well as the
lowering of $\gamma_s$. If we had used the older values
$\Xi^-+\bar{\Xi}^+=8.19\pm1.06$, $\Xi^-=7.23\pm0.88$ [15] and
$\bar{\Lambda}/\Lambda=0.20\pm0.04$ [16] we would get $T=172.3\pm9.3$ MeV,
$\mu_u=70.4\pm7.4$ MeV, $\mu_d=80.3\pm7.8$ MeV, $\mu_s=10.4\pm9.7$ MeV
and $\gamma_s=0.827\pm0.088$ with $\chi^2/dof=22.8/7$ and even larger values
for $T$ and $\gamma_s$ if the pions were excluded. These results are compatible
with the results of Analysis B for the $Pb+Pb$ case in [16].

From Fig. 1 it is evident that, the source of the produced multiparticle
system in the $Pb+Pb$ interaction at 158 AGeV lies well within the hadronic
domain, which equivalently means that the deconfined, partonic phase has not
been attained in this system.

It is evident when comparing the last columns of Table 2 that neither a
notable improvement of the fit (e.g. the value of $\chi^2/dof$) nor a
remarkable change in the fitted thermodynamic variables occurs when the
$<\pi>$ multiplicity (which contains pions) is excluded from the data, as is
the case with the fits in the $S+S$ [10] and $S+Ag$ [11] data.
The same conclusion can be inferred from Fig. 2, where the fittings of
particle multiplicities in $4\pi$ phase space,
with and without the inclusion of the pion multiplicities, are compared to
the experimental values. The overall picture is that there is no notable
difference between the two cases, meaning that the produced entropy (mainly
associated with the pion production) is well accounted for by the hadronic
thermal model.

A summary of results attained through the SSBM analysis of $S+S$ [10] and
$S+Ag$ [11] interactions at 200 AGeV (NA35), as well as the $p+\bar{p}$ (UA5) at
several energies [10], is exhibited in Fig. 3. It should be pointed out that
the surface setting boundaries on the hadronic domain is a constraint among
the thermodynamic variables $(T,\lambda_u,\lambda_d,\lambda_s,\gamma_s)$. The
projection of this surface on the $(T,\mu_B)$ plane which corresponds to the
particular value of $\gamma_s$ resulting from the fit to the experimental
multiplicities is plotted for every interaction. Thus every point of
interaction should be compared with the relevant projection.
The analysis shows that the $S+S$
source is situated mostly outside the hadronic phase
(with probability of being outside $74\%$), whilst the
$S+Ag$ is located just beyond the deconfinement line
(with probability of being outside $52\%$)\footnote{It should be
noted that, as far as experiments whose multiparticle data flirt with the
critical curve, special considerations of computational nature are called
for. Such issues have been discussed at length in Refs [10,11]. Moreover, as an
additional aid to probing the region beyond the critical surface we conducted
relevant analyses for an increased value of $T_0$.}. In addition, the
analysis exhibits a large ($\sim30\%$) entropy enhancement of the experimental
negative-hadron yields compared to the model, an effect observed also by other
calculations [1-3]. This enhancement may be attributed to contributions
from the deconfined quark phase with many libarated new partonic degrees of
freedom. Also, the thermal-statistical models [1-3], which do not posses
inherently any bounds for the hadronic phase, give fitted source temperature
beyond those of the corresponding SSBM analysis: $T\simeq 180-200$ MeV, which
is about $5-25$ MeV higher.

We conclude that all these results are corroborating our position that the
$Pb+Pb$ interaction at the maximum energy of 158 AGeV is well within the
hadronic phase. The $S$-induced interactions and in particular the $S+S$ one
are located beyond the HG phase, in the deconfined quark-gluon domain. For
the $p+\bar{p}$ collision at $\sqrt{s}=200$ to $900$ GeV, the SSBM analysis
clearly locates this system within the hadronic domain\footnote{The
$p+\bar{p}$ thermodynamic variables have been obtained through a
grand-canonical analysis. It is known that canonical suppression is
relevant to the $p+\bar{p}$ interaction and it may affect the extraction of
parameters. It cannot, however, move the point outside the critical surface
since it lies well inside the hadronic domain.}.

In summary, the potential of SSBM to perform not only thermal-statistical
fits to multiparticle systems, but also to define the limits of the hadronic
phase has been employed to analyze and asses several nucleus-nucleus
interactions. On the basis of these analyses we reach the conclusion that the
$Pb+Pb$ interaction at 158 AGeV has not crossed the deconfinement line,
remaining within the hadronic phase. On the other hand, the $S+S$ interaction
at 200 AGeV appears to have produced a deconfined partonic state, for the
first time.

\newpage
\vspace{0.3cm}
{\large{\bf Table Captions}} 
\newtheorem{f}{Table} 
\begin{f} 
\rm Experimentally measured full phase space multiplicities in the 
NA49 $Pb+Pb$ experiment at 158 AGeV and their theoretically fitted values by SSBM, 
with the inclusion of the $<\pi>$ multiplicity and without it.
\end{f}
\begin{f} 
\rm Results of the analysis by SSBM of the experimental 
data from $Pb+Pb$ experiment ($4\pi$ phase space), with the inclusion of the 
$<\pi>$ multiplicity and without it. 
\end{f}
\begin{f}
\rm Particle ratios from the experimentally measured full phase space 
multiplicities in the $Pb+Pb$ experiment at 158 AGeV, used in the analysis. 
\end{f}

\begin{center}

{\bf $\bf Pb+Pb$ (NA49) Full phase space}
\begin{tabular}{|c|ccc|} \hline 
\hspace{0.3cm}
Particles           & Experimental  & Calculated   & Calculated      \\ 
                    & Data          & with $<\pi>$ & without $<\pi>$ \\
\hline\hline
$<\pi>^{\rm a}$     & $600\pm 30$   & 570.51$^b$   & 527.05$^c$ \\
$K^+$               & $95 \pm 10$   & 96.583       & 93.470     \\
$K^-$               & $50 \pm 5$    & 58.168       & 54.528     \\
${K_s}^0$           & $60\pm 12$    & 76.110       & 72.800     \\
$p$                 & $140\pm 12$   & 150.73       & 148.01     \\
$\overline{p}$      & $10\pm 1.7$   & 8.1956       & 7.2435     \\ 
$\phi$              & $7.6\pm 1.1$  & 7.4101       & 7.3914     \\
$\Xi^-$             & $4.42\pm 0.31$& 4.0838       & 4.1886     \\ 
$\overline{\Xi}^+$  & $0.74\pm 0.04$& 0.76299      & 0.77076    \\
$B-\overline{B}$    & $362\pm 12$   & 366.65       & 364.43     \\
$\overline{\Lambda}$& $5.14 \pm 0.6$& 4.8591       & 4.6047     \\ 
$\Lambda$           & $52\pm 3$     & 48.263       & 48.472     \\ \hline 
\end{tabular} 
\end{center} 

{\footnotesize
$^{\rm a}$ $<\pi>\equiv (\pi^+ + \pi^-)/2$

$^b$ A correction factor 1.03272 has been included for the effect of
Bose statistics.

$^c$ A correction factor 1.03107 has been included for the effect of Bose 
statistics. This multiplicity is not included in the fit. }

\begin{center} 
Table 1 
\end{center}

\begin{center}
{\bf $\bf Pb+Pb$ (NA49) Full phase space} 
\begin{tabular}{|c|cc|} \hline 
Fitted Parameters & Fitted with $<\pi>$ & Fitted without $<\pi>$ \\
\hline\hline
$T$ (MeV)         & $156.3 \pm 4.2  $    & $157.4 \pm 4.3$   \\
$\lambda_u$       & $1.606 \pm 0.051$    & $1.633 \pm 0.064$ \\ 
$\lambda_d$       & $1.686 \pm 0.059$    & $1.721 \pm 0.077$ \\ 
$\lambda_s$       & $1.178 \pm 0.030$    & $1.171 \pm 0.031$ \\
$\gamma_s$        & $0.758 \pm 0.068$    & $0.790 \pm 0.081$ \\
$VT^3/4\pi^3$     & $15.7  \pm   2.8$    & $13.8  \pm 3.4$   \\ 
$\chi^2/dof$      & $10.85\;/\;8$        & $8.46\;/\;7$      \\ 
$\mu_u$ (MeV)     & $74.0  \pm   5.3$    & $77.2  \pm 6.6$   \\ 
$\mu_d$ (MeV)     & $81.7  \pm   5.9$    & $85.5  \pm 7.5$   \\ 
$\mu_s$ (MeV)     & $25.6  \pm   4.0$    & $24.9  \pm 4.2$   \\
\hline 
\end{tabular} 
\end{center} 

\begin{center} 
Table 2
\end {center}

\begin{center}
{\bf $\bf Pb+Pb$ (NA49) Full phase space} 
 
\begin{tabular}{|c|c|} \hline 
Particle Ratios used                 & Experimental   \\ 
for the plots                        & Values         \\ 
\hline\hline 
$<\pi>/(B-\overline{B})$             & $1.66   \pm  0.10  $ \\
$K^+/(B-\overline{B})$               & $0.262  \pm  0.029 $ \\
$K^-/(B-\overline{B})$               & $0.138  \pm  0.015 $ \\
${K_s}^0/(B-\overline{B})$           & $0.166  \pm  0.034 $ \\
$p/(B-\overline{B})$                 & $0.387  \pm  0.036 $ \\
$\overline{p}/(B-\overline{B})$      & $0.0276 \pm  0.0048$ \\ 
$\phi/(B-\overline{B})$              & $0.0210 \pm  0.0031$ \\
$\Xi^-/(B-\overline{B})$             & $0.0122 \pm  0.0009$ \\ 
$\overline{\Xi}^+/(B-\overline{B})$  & $0.00204\pm 0.00013$ \\
$\overline{\Lambda}/(B-\overline{B})$& $0.0142 \pm  0.0017$ \\ 
$\Lambda/(B-\overline{B})$           & $0.144  \pm   0.010$ \\ \hline 
\end{tabular} 
\end{center} 
                                                
\begin{center} 
Table 3 
\end{center}

\newpage
\vspace{0.3cm}
{\large{\bf Figure Captions}} 
\newtheorem{g}{Figure} 
\begin{g} 
\rm Experimental particle ratios in the $(\mu_u,T)$-plane for the Pb+Pb
interaction measured in $4\pi$ phase space with $\gamma_s$ set to 0.76.
The point and the cross correspond to the $\chi^2$ fit with the $<\pi>$. The
thick solid line represent the limits of the hadronic phase (HG) as set by
the SSBM. The smallest experimental value of the ratio $p/(B-\bar{B})=0.351$
cannot be depicted on the $(\mu_u-T)$ plane because it has no solution  for the
given variables. Thus the space which is compatible with the experimental
values is the one which is enclosed by the ratio with the largest experimental
value.

\end{g}
\begin{g}
\rm Comparison between the experimentally measured multiplicities in $4\pi$
phase space and the theoretically calculated values in the fit with $<\pi>$
and without $<\pi>$ for the $Pb+Pb$ interaction. The difference is measured
in units of the relevant experimental error.
\end{g}
\begin{g}
\rm $(\mu_B,T)$-phase diagramme with points obtained from fits to $p+\bar{p}$,
$S+S$, $S+Ag$ and $Pb+Pb$ data and corresponding critical curves given by
SSBM.
\end{g}

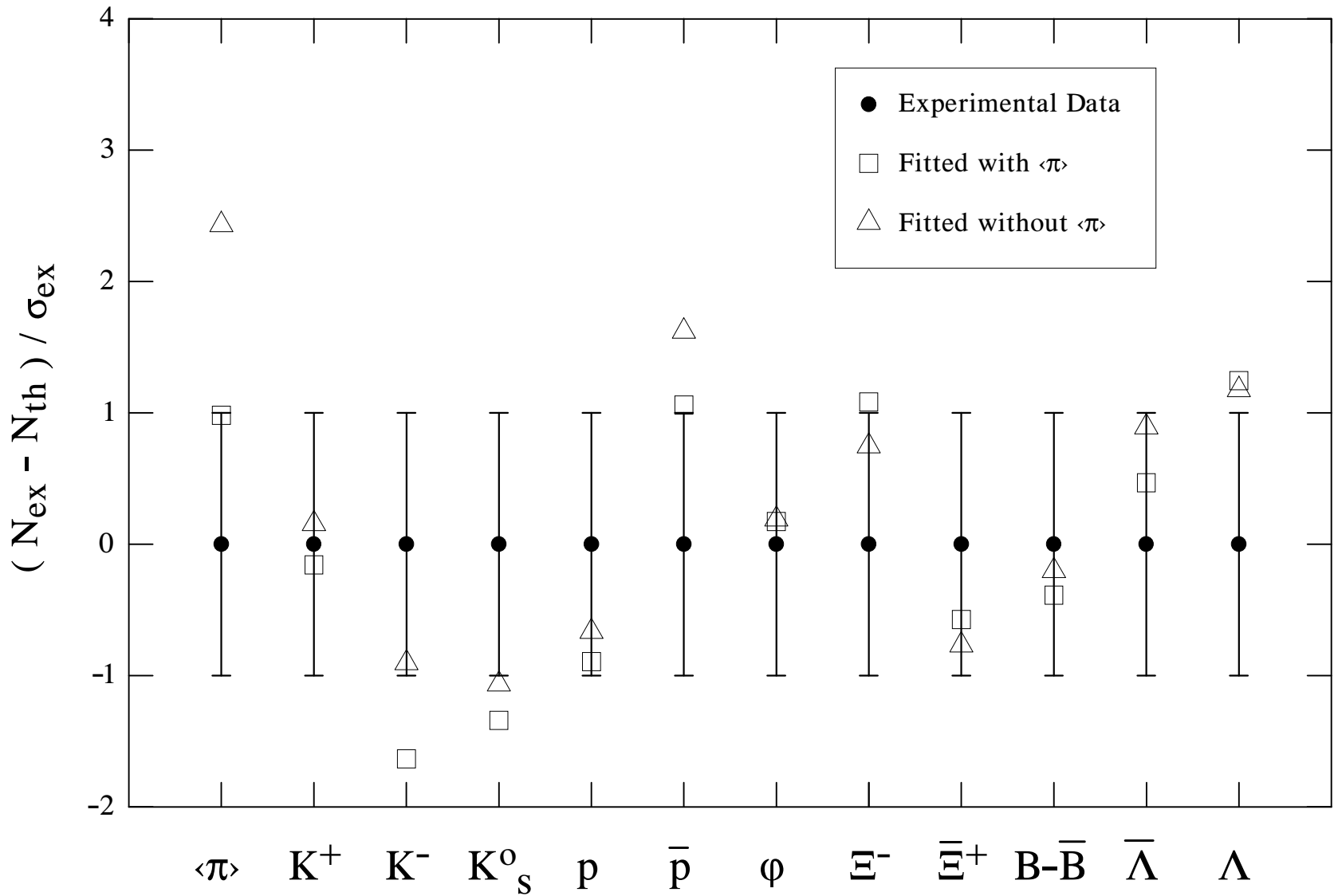
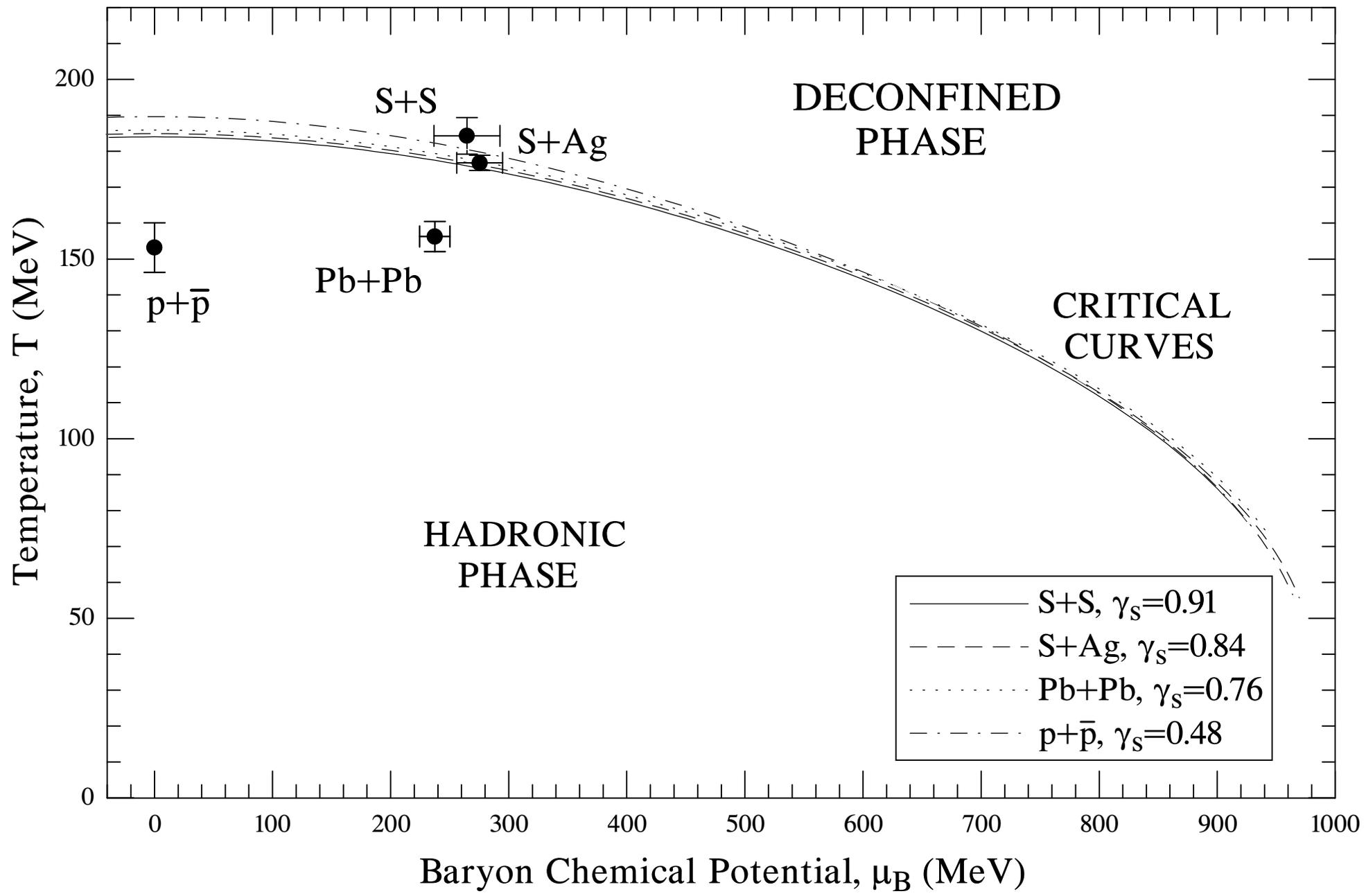
\end{document}